\documentclass[12pt]{article}
\usepackage[english]{babel}
\usepackage{psfig}

\begin{document}

\title{Time evolution of wave-packets in quasi-1D disordered media}

\author{A. Politi$^1$, S. Ruffo$^2$, L. Tessieri$^3$
\bigskip \\
{\it $^1$ Istituto Nazionale di Ottica 50125 Firenze} \\
{\it Italy and INFM-Firenze ; e-mail {\tt politi@ino.it}} \\
{\it $^2$ Dipartimento di Energetica ``S. Stecco",} \\ 
{\it Universit\`a di Firenze, via s. Marta 3 50139 Firenze Italy,} \\
{\it INFN and INFM; e-mail {\tt ruffo@avanzi.de.unifi.it}} \\
{\it $^3$ Instituto de F\'{\i}sica, Universidad Aut\'{o}noma de Puebla,} \\
{\it Apdo. Postal J-48, Puebla, Pue. 72570, M\'{e}xico;} \\ 
{\it e-mail {\tt tessieri@sirio.ifuap.buap.mx}}}

\date{27th October 1999}
\maketitle

\begin{abstract}
We have investigated numerically the quantum evolution of a $\delta$-like 
wave-packet in a quenched disordered medium described by a tight-binding 
Hamiltonian with long-range hopping (band random matrix approach).
We have obtained clean data for the scaling properties in time 
and in the bandwidth $b$ of the packet width $\tilde{M}$ and its 
fluctuations $\Delta_{\tilde{M}}$ with respect to disorder realizations. 
We confirm that the fluctuations of the packet width in the steady-state
show an anomalous scaling $\Delta_{\tilde M} /\tilde{M} \sim b^{-\delta}$ 
with $\delta = 0.75 \pm 0.03$. This can be related to the presence of
non-Gaussian tails in the distribution of $\tilde{M}$. 
Finally, we have analysed the steady state probability profile and we have
found $1/b$ corrections with respect to the theoretical formula derived
by Zhirov in the $b \to \infty$ limit, except at the origin, where the
corrections are $O(1/\sqrt{b})$.

\end{abstract}


\section{Introduction}

Band random matrices (BRM) represent an effective model for both
1D disordered systems with long-range hopping and
quasi-1D wires\cite{Guh98}. The bandwidth $b$ plays the role of the range
of the interaction in the first case, the one of the square
root of the number of independent conduction channels in the second.
Up to now, studies have been mostly devoted to the analysis of
the stationary solutions of the Schr\"odinger equation and to the corresponding
spectral properties of BRM's~\cite{Cas90}. Much less
is known about the solutions of the time dependent Schr\"odinger equation,
a topic on which only a few studies have been performed~\cite{Izr96,Izr97}.
The partial analogy of this latter problem with the 
`dynamical localization' phenomenon in the kicked rotor~\cite{Casa90} 
suggests that an initial delta-like packet spreads diffusively and
eventually saturates to a localized state. The width of this
asymptotic packet for BRM's is of the order of $b^2$ lattice sites, i.e.
the same order as the localization length of all the 
eigenfunctions~\cite{Cas90}.

The theoretically predicted scaling laws for the mean square displacement
$\tilde{M}$ were tested numerically and a comparison of the asymptotic
form of the wave-packet with a theoretical formula~\cite{Gogo76} derived 
for the 1D Anderson model was attempted~\cite{Izr97}.
More recently some new theoretical results appeared which give
a formula for the time asymptotic packet in the BRM model
in the large $b$ limit~\cite{Zhi97}.
Therefore, it became important to check numerically this formula and
to both investigate how the packet reaches its time asymptotic 
shape and measure the size of finite $b$ corrections.  For what
the time evolution is concerned some phenomenological expressions
were suggested in ref.~\cite{Izr97}, based on a power-law convergence 
of the mean square displacement to its steady state value. However, the 
presence of large statistical fluctuations prevented the authors of
ref.~\cite{Izr97} from assessing whether the time asymptotic scaling is
ruled by power law corrections or by the logarithmic corrections to the
$t^{-1}$ dependence suggested by rigorous results obtained for the 1D
Anderson model~\cite{NPF87}.
The fluctuations $\Delta_{\tilde{M}}$ of the width of the asymptotic 
wave-packet with the realization of the disorder constitute an even 
more controversial issue,
since not even the scaling behaviour is clearly understood. Some evidence 
of an anomalous behaviour was presented in two previous studies of the 
same problem~\cite{Izr96,Izr97} and in the kicked 
rotor~\cite{Cas94}. In all cases the numerics was too poor to make 
a convincing statement about the value of the anomalous exponent. 

The bottleneck of the previous simulations was the slowness of the integration
scheme, a 4-th order Runge-Kutta with a small time step to obtain
a good conservation of probability over a long time span. This low
efficiency prevented from reaching sufficiently 
large values of $b$ and from considering a large enough number of 
realizations of the BRM's. We have instead implemented a 2-nd order Cayley 
algorithm, which, being unitary, exactly conserves probability,
although the one-step integration error is larger than the one of the
Runge-Kutta scheme (a situation similar to those of symplectic
algorithms in classical Hamiltonian dynamics). This has allowed us to more 
than double the maximum bandwidth (from $b=12$ to $b=30$) and to 
increase the statistics by a factor four (in the worst case).

As a result, we have been able to complete an accurate analysis
of the time evolution of the mean square displacement, finding 
that there is no need to invoke effective formulas with a power-law 
time dependence even at relatively short times.
We have found a clean evidence of an anomalous scaling of the
relative fluctuations of the packet width, which behave as
$\Delta_{\tilde{M}} /\tilde{M} \sim b^{-\delta}$ with 
$\delta = 0.75 \pm 0.03$. 
In order to confirm this anomaly, we have investigated the
statistics of the packet width at a specific time in the localization regime. 
The probability distributions at various $b$ values, when appropriately
rescaled, superpose, and the resulting universal ($b$ independent) curve 
is definitely different from a Gaussian with an exponential tail 
at large $\tilde{M}$ values.
Finally, we have compared our results with the theoretical 
formula for the asymptotic wave-packet~\cite{Zhi97} finding a convincing
agreement.
The finite $b$ corrections to the $b \to \infty$ Zhirov expression 
are of order $(1/b)$.

\section{Model and numerical technique}

We have considered the time-dependent Schr\"{o}dinger equation
\begin{equation}
i \frac{\partial \psi_{i}}{\partial t} = \sum_{j=i-b}^{i+b} H_{ij}
\psi_{j} 
\label{schroe}
\end{equation}
where $\psi_{i}$ is the probability amplitude at site $i$ and
the tight-binding Hamiltonian $H_{ij}$ is a real symmetric band random
matrix. The band structure of the Hamiltonian is determined by the
condition
\begin{displaymath}
\begin{array}{ll}
H_{ij}=0 & \mbox{if $|i-j| > b$},
\end{array}
\end{displaymath}
the parameter $b$ setting the band-width; the matrix elements inside
the band are independent Gaussian random variables with 
\begin{displaymath}
\begin{array}{lll}
\langle H_{ij} \rangle_d = 0 & \mbox{and} & \langle \left( H_{ij}
\right)^{2} \rangle_d = 1 + \delta_{ij}
\end{array}
\end{displaymath}
where the symbol $\langle \cdot \rangle_d$ stands for the average over
different realizations of the disorder.
In the present work we have considered the evolution of an `electron'
initially localized at the centre (identified with the site $i=0$) of an
infinite lattice. To this aim we have analysed the solution of
equation~(\ref{schroe}) corresponding to the initial condition
\begin{displaymath}
\psi_{i} \left( t=0 \right) = \delta_{i0} .
\end{displaymath}

Since the wave-packet evolves in a supposedly infinite lattice, it is 
necessary to avoid any spurious boundary effect due to the inevitably
finite size of the vectors used in the numerical computations. This goal 
has been achieved by resorting to a self-expanding lattice, i.e. a lattice 
whose size is progressively enlarged according to the development of 
the wave-function. At each integration step, our program 
checks the probability that the electron is in the leftmost and rightmost 
$b$ sites, adding $10b$ new sites whenever the amplitude
$|\psi_i|$ is larger than $\varepsilon=10^{-3}$ in at least one of the $2b$
outermost sites.
We have separately verified that $\varepsilon$ is small enough not to
significantly affect the computation of the probability distribution.
For instance, by lowering $\varepsilon$ by an order of magnitude, the mean
squared displacement (computed over the same disorder realizations) changes
only by a few percent. Since this systematic error is not larger than the 
uncertainty due to statistical fluctuations, it is not convenient to
reduce the cut-off as it would turn out in a slower code with a consequent
reduction of the statistics.

The Schr\"{o}dinger equation~(\ref{schroe}) was integrated by 
approximating the evolution operator $\exp \left( {-iHt} \right)$
with the Cayley form 
\begin{equation}
\exp \left( -i H \delta t \right) \simeq \frac{1 - i H \delta t/2}{1 + i
H \delta t/2} ,
\label{cayley}
\end{equation}
which implies that the values of the wave-function at two successive
time-steps are related by
\begin{equation}
\left( 1 + \frac{1}{2} i H \delta t \right) \psi(t+\delta t) =
\left( 1 - \frac{1}{2} i H \delta t \right) \psi(t) .
\label{cayley2}
\end{equation}
Solving the band diagonal system of equations~(\ref{cayley2}) allows one
to determine $\psi(t+\delta t)$ once $\psi(t)$ is known.
Cayley's algorithm is a standard tool for the computation of the solutions
of the Schr\"{o}dinger equation (see for instance ref.~\cite{Pre92}); to the
best of our knowledge, this is the first application to the specific field of
random Hamiltonians with long-range hopping.
Cayley's form~(\ref{cayley}) for the evolution operator has two relevant
features: it is second-order accurate in time and unitary;
in addition, the corresponding integration scheme~(\ref{cayley2}) is
stable. Stability is essential in order to study the long time
evolution of the wave-packet; as for unitarity, it ensures the
conservation of probability and, together with second-order accuracy in
time, allows one to choose time steps $\delta t$ two or three order of
magnitude bigger than those used in Runge-Kutta integration schemes.
Indeed, we could make use of a time step $\delta t \sim 10^{-1}$, to be
compared with the time step $\delta t \sim 10^{-4} - 10^{-3}$ used for
the same problem in Refs.~\cite{Izr96,Izr97}.
To ascertain how large a $\delta t$ could be used, we have compared 
the solutions obtained through Cayley's algorithm at various $\delta t$ 
with the exact solution of the Schr\"{o}dinger equation~(\ref{schroe}), 
computed by diagonalizing
the Hamiltonian (to avoid boundary effects due to the finite size of the
diagonalized matrices, we have considered sufficiently short evolution
times).
By this way we came to the somewhat surprising conclusion that the
validity range of the approximate equality~(\ref{cayley}) extended up to
time steps as big as $\delta t \sim 1/\sqrt{b}$ (the scaling of $\delta
t$ with the band-width $b$ was necessary to compensate the opposite
scaling of the energy eigenvalues with $\sqrt{b}$).
To check this conclusion, we have computed the mean squared displacement 
in the localized regime for several values of $\delta t$ in the range
$10^{-2}/\sqrt{b} - 1/\sqrt{b}$, finding differences of a few percent, 
not larger than the statistical fluctuations.

This can depend on the fact that the long time evolution of the
wave-packet seems to be led by the eigenstates at the band centre.
Indeed, in the energy representation, the exact evolution operator and the 
Cayley form can be written as
\begin{eqnarray*}
\exp \left(-i H \delta t \right) = \sum_{n} |n \rangle e^{-i E_{n}
\delta t} \langle n | \\ 
\frac{1 - i H \delta t/2}{1 + i H \delta t/2}
 = \sum_{n} |n \rangle e^{-i \phi_{n}(\delta t) } \langle n |~,
\end{eqnarray*}
with
\begin{displaymath}
\phi_{n}(\delta t) = 2 \arctan \left( E_{n} \delta t / 2 \right)~,
\end{displaymath}
where $|n\rangle$ is the eigenvector corresponding to the energy $E_{n}$.
These equations show that, for increasing $\delta t$, the
approximate equality~(\ref{cayley}) holds true only in the subspace
spanned by the eigenvectors corresponding to the band centre, while at
the band edge, where the eigenvalues $|E_{n}|$ tend to $\sqrt{b}$, the
coefficients $\exp \left( -i E_{n} \delta t \right)$ and $\exp \left( -i
\phi_{n} \delta t \right)$ become quickly different. Therefore, the
eigenstates at the band edges appear to play a minor role in the
time evolution. This is probably due to the shorter localization length
of such states compared with those in the centre: in fact, for the mean
square displacement, all eigenstates are weighted with their localization
length~\cite{xxx}.

\section{Results}

To investigate the time evolution of the wave-packet, we have
computed the mean square displacement
\begin{equation}  
\label{MSD}
\tilde M(b,t)=\left\langle u(t)\right\rangle \equiv \left\langle
\sum_{j=-\infty}^{\infty} j^2 |\psi_j(t)|^2\right\rangle_d \,\, .
\end{equation}

Previous studies of this problem strongly suggest that
$\tilde M$ satisfies the scaling relation
\begin{equation}  
\label{scal}
\tilde M(b,t) = b^4 M (\tau =t/b^{3/2}) \quad ,
\end{equation}
for large enough values of the bandwidth $b$. Nevertheless, in 
Ref.~\cite{Izr97}, where the most detailed numerical investigation has
been carried out, it was not possible to obtain a clear verification
of the scaling law~(\ref{scal}) due to the poor statistics and the 
small values of $b$.
 
\begin{figure}[hbt]
\begin{center}
\psfig{figure=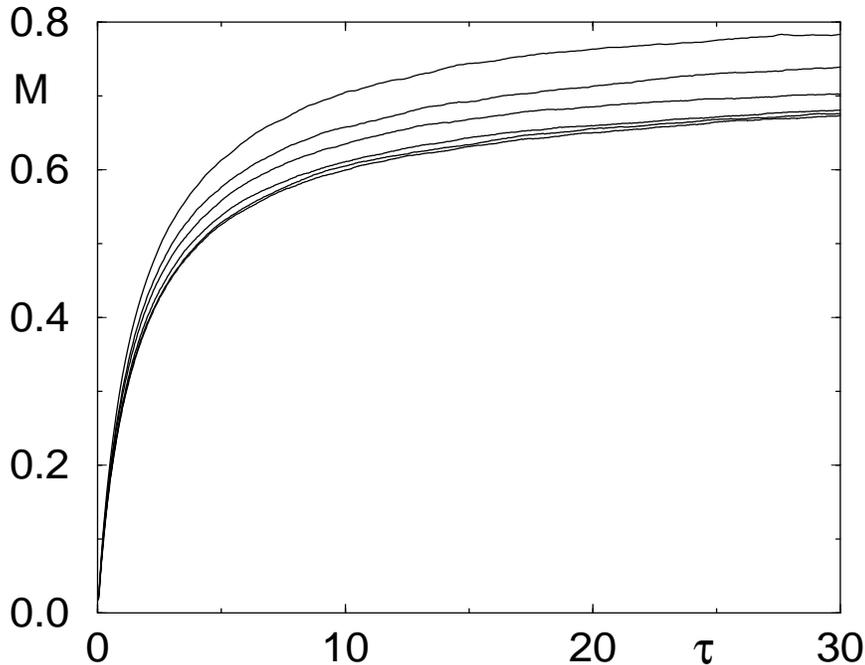,height=3.5in,width=4.5in}
\caption{Rescaled mean squared displacement $M$ vs. rescaled
time $\tau=t/b^{3/2}$. From top to bottom: $b = 8,12,16,22,26,30$}
\label{fig1}
\end{center}
\end{figure}

The faster integration algorithm described in the previous section has
allowed us both to average over more realizations (400 in the worst case), 
i.e to reduce statistical fluctuations, and to reach 
larger values of $b$ (namely, $b=30$ instead of $b=12$ as in
Ref.~\cite{Izr97}). The results reported in Fig.~\ref{fig1}
for several values of $b$ are clean enough to show a convincing
convergence from above to a limit shape. In other words, there is no
possibility to interpret the deviations as a signature of a different
scaling behaviour for $\tilde M$.

\begin{figure}[hbt]
\begin{center}
\psfig{figure=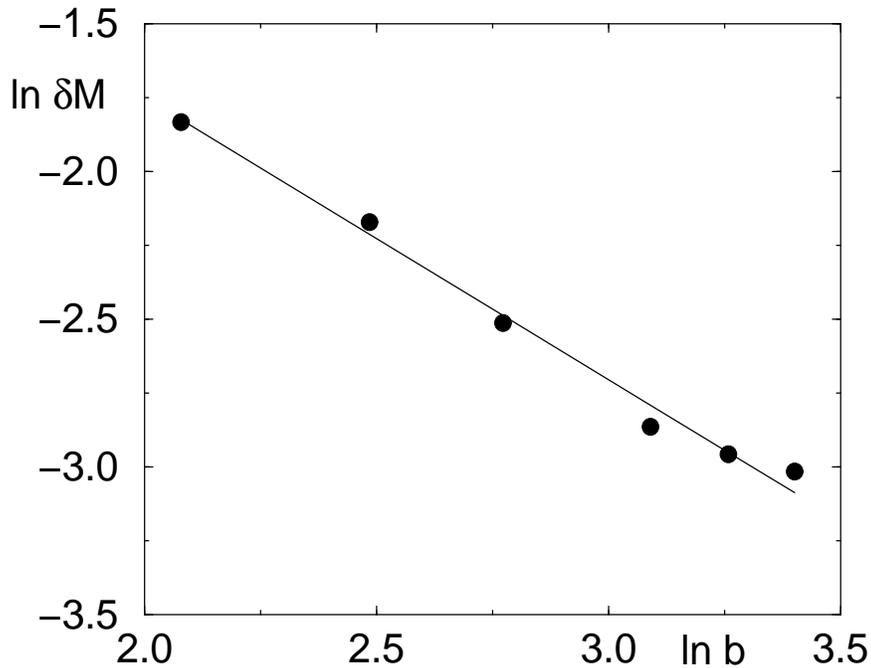,height=3.5in,width=4.5in}
\caption{Convergence towards zero of the finite band
correction (\ref{devia}). The slope is 0.95 thus close to -1.}
\label{fig2}
\end{center}
\end{figure}

In order to perform a more quantitative analysis, we have proceeded
in the following way: $M(\tau,b)$\footnote{We have added the variable $b$
to underline the residual but asymptotically irrelevant dependence on the 
band-with.} has been averaged over 
the time interval $20<\tau<30$ to obtain the more statistically reliable 
quantity $\langle M \rangle_t (b)$. By assuming a dependence of the type 
\begin{equation}  
\label{fit1}
\langle M \rangle_t (b) = M_\infty ( 1 + a b^{-\alpha}) ,
\end{equation}
we have fitted the three parameters $M_\infty$, $a$ and
$\alpha$, finding that the convergence rate $\alpha$ is very close to 1 
(0.95), i.e. that the finite-band corrections are of the order $1/b$.
The fitted value of $M_\infty$ is 0.61. The results for the finite-band 
correction
\begin{equation}
\delta M = M_\infty - \langle M \rangle_t(b)
\label{devia}
\end{equation}
are plotted in Fig.~\ref{fig2}.

The good quality of our numerical data suggests also the possibility to 
compare the temporal behaviour with the available theoretical formulas. 
In particular, it has been argued in Ref.~\cite{NPF87} that the existence 
of the so-called Mott states should imply a $(\ln t)/t$ convergence of 
$M$ to its asymptotic value. Therefore, we propose the following
expression
\begin{equation}
\label{fit2}
M(\tau,b) = M(\infty,b) \left( 1- \frac{1 + A \ln (1 + \tau/{t_D})}
   {1+\tau/{t_D}} \right) \; ,
\end{equation}
which is the simplest formula that we have found able to 
reproduce also the initially linear (i.e. diffusive) regime.  For each 
value of $b$, the best fit is so close to the numerical data of
Fig.\ref{fig1} to be almost indistinguishable from them (this is
why we do not report the fits on the same figure).
The meaningfulness of the above expression is further strengthened by the
stability of the three free parameters $M(\infty,b)$, $A$ and $t_D$,
which allows the calculation of a $b$-independent diffusion
constant (see below).
From the values of $M(\infty,b)$, we can extrapolate the asymptotic 
value $M(\infty,\infty)$ in exactly the same way as we have done 
for $\langle M \rangle_t(b)$, finding once more a $(1/b)$-convergence to 
a value around 0.70. This result is to be compared
with the theoretical prediction $M(\infty,\infty) \approx
0.668$~\cite{Zhirov}.
The deviation of about 0.03 can be attributed to the accuracy of
the integration algorithm.

Another important parameter that can be extracted from 
formula~(\ref{fit2}) is the diffusion coefficient. Indeed, by expanding
Eq.~(\ref{fit2}) for small $\tau$, we find 
\begin{equation}
\label{diff}
\frac{M(\tau,b)}{\tau} = \frac{M(\infty,b)}{t_D}(1-A) = D 
\end{equation}
The diffusion constant $D$ turns out to be close to 0.50 for all values of
$b$ and, what is more important, close to the value that we obtain
from a quadratic fit of the initial growth rate of the packet. This is
a very encouraging result, since it confirms the correctness of
formula (\ref{fit2}) for both the diffusive and the localized
regimes. Let us notice that the value $D \simeq 0.50$ is somewhat 
smaller than the one reported in~\cite{Izr97} ($D \simeq 0.83$). Taking
into account statistical fluctuations and systematic deviations, we find 
that $D = 0.50 \pm 0.05$.

In past papers, a phenomenological expression involving a power-law
convergence to the asymptotic value of the mean squared displacement
has been proposed~\cite{Izr97,I90}, arguing that it should provide an
effective description of both the diffusive and localized regime
\begin{equation}
\label{fit3}
M(\tau,b) = M(\infty,b) \left( 1-\frac 1{\left( 1+\tau/{t_D}\right) ^\beta }
\right) \; .
\end{equation}
The success of expression (\ref{fit2}) shows that there is no need to 
introduce anomalous power laws to reproduce the numerical findings. However, 
for the sake of completeness, we have fitted our numerical data also with 
Eq.~(\ref{fit3}), finding an equally good agreement. Therefore, on the basis
of the quality of the fit we cannot conclude which of the two expressions is 
better; nevertheless, it is worth recalling that the former one has the 
correct asymptotic behaviour and, moreover, the fitted parameters are 
more stable.

\begin{figure}[hbt]
\begin{center}
\psfig{figure=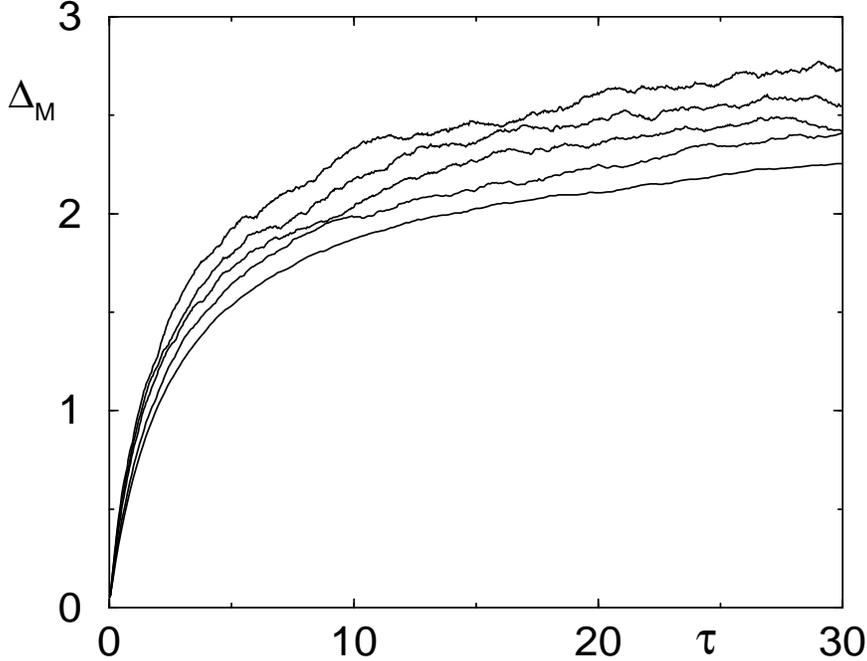,height=3.5in,width=4.5in}
\caption{Rescaled fluctuations of the mean squared displacement
$\Delta_M$ in formula (\ref{scal2}) with $\nu=3$ vs. rescaled time
$\tau=t/b^{3/2}$. The values of $b$ increase from bottom to top
$b=8,12,22,26,30$. There is a clear tendency to grow for increasing
$b$ values.}
\label{fig3}
\end{center}
\end{figure}

A much more controversial situation exists about the fluctuations of the
packet width. Let us introduce the r.m.s. deviation
\begin{equation}
\Delta_{\tilde{M}}(b,t)\equiv \sqrt{ \left\langle u(t)^2\right\rangle_d
\,\,-\,\left\langle u(t)\right\rangle_d^2}~.  
\label{DM}
\end{equation}
In fact, it has not yet been clarified how the above variable scales 
in the large-$b$ limit. In particular, the correct value of 
the scaling exponent $\nu$ in the relation 
\begin{equation}  
\label{scal2}
\Delta_{\tilde{M}}(b,t) = b^\nu \Delta_M (\tau =t/b^{3/2},b) \quad
\end{equation}
is still unknown; this is why the $b$ dependence in $\Delta_M$
is explicitly maintained.
A scaling like $b^4$ would imply that the packet-width is not a 
self-averaging quantity, since the relative size of the fluctuations would 
not go to zero for increasing $b$. Conversely, an exponent $\nu=3$ 
corresponds both to self-averaging and `normal' behaviour. 
In fact, the number $N_c$ of independent channels (lattice sites) actively 
contributing to the localized region (i.e. the localization length) is of 
the order $b^2$. If we assume that all such contributions to the second 
moment $\tilde{M}$ are independent of one another, then we are led to
conclude that 
the relative fluctuations should decrease as $1/\sqrt{N_c} = 1/b$, thus 
yielding an absolute growth as $b^3$. Since previous studies~\cite{Izr97} have 
suggested a small anomaly, i.e. $\nu$  slightly larger than 3, we have
chosen to report the behaviour of $\Delta_M(\tau,b)$ for $\nu=3$.
The data shown in Fig.~\ref{fig3} reveals a drastically different 
behaviour from what observed in Fig.~\ref{fig1}. First of all, the curves 
tend to grow for increasing $b$; moreover, there is no obvious indication 
of a convergence to some finite value. Altogether, these features imply 
that $\nu$ is strictly larger than 3, qualitatively in agreement with 
previous simulations. 

\begin{figure}[hbt]
\begin{center}
\psfig{figure=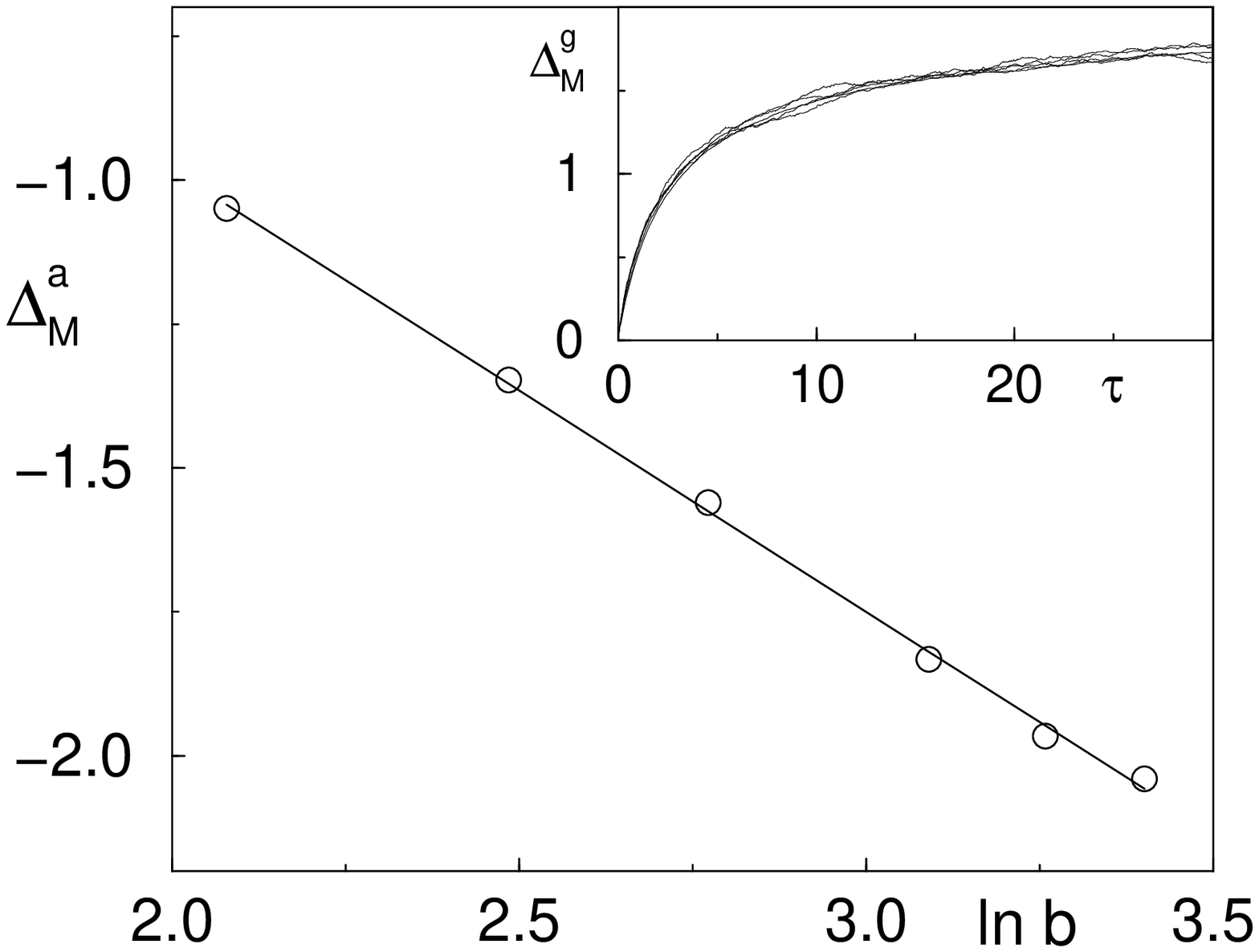,height=3.5in,width=4.5in}
\caption{Averaged and normalized fluctuations of the mean
squared displacement in formula (\ref{scal3}) vs. $b$ (log-log scale).
The line is a fit with a power law $b^{-\delta}$ with $\delta=0.75$.
In the inset we report $\Delta_M^g(\tau)$ (see Eq.~(\ref{scal4}) for its
definition) for the same values of $b$ as in Fig.~\ref{fig3} and 
$\delta = 0.75$.}
\label{fig4}
\end{center}
\end{figure}

In order to perform a more quantitative analysis, we have computed
the average of $\Delta_{\tilde{M}}$ and rescaled it to 
the average of $\tilde M$, 
\begin{equation}
\label{scal3}
{\Delta_M^a}(b) \equiv \frac{\langle \Delta_{\tilde{M}}\rangle_t}
   {\langle \tilde M \rangle_t} \; .
\end{equation}
($\langle \cdot \rangle_t$ is again to be interpreted as the
average over the time interval $20<\tau<30$).
The advantage of this renormalization, already adopted in Ref.~\cite{Izr97}, 
is that it reduces finite-band corrections. The results reported 
in Fig.~\ref{fig4},
reveal a clean power law decay with an exponent $\delta \approx 0.75$. 
This value is slightly larger than the one found in the previous studies, 
but follows from a much cleaner numerics. A more global check of the
scaling behaviour can be made by plotting the rescaled fluctuations
\begin{equation}
\label{scal4}
{\Delta_M^g}(\tau) \equiv b^\delta \, \frac{\Delta_{\tilde{M}}}{\tilde M} \; .
\end{equation}
for the various values of $b$. The optimal value of $\delta$ can thus be 
identified as that one yielding the best data collapse. The curves reported
in the inset of Fig.~\ref{fig4} have been obtained for $\delta = 0.75$. 
It is necessary to modify $\delta$ by at least $\pm 0.03$ units in order
to see a significant worsening of the data collapse. Accordingly, the 
best estimate of the anomalous exponent is $\delta = 0.75 \pm 0.03$, so 
that the dependence of $\Delta_M$ on $b$ in Eq.~(\ref{scal2}) is removed for
$\nu=4-\delta\approx 3.25$.

\begin{figure}[hbt]
\begin{center}
\psfig{figure=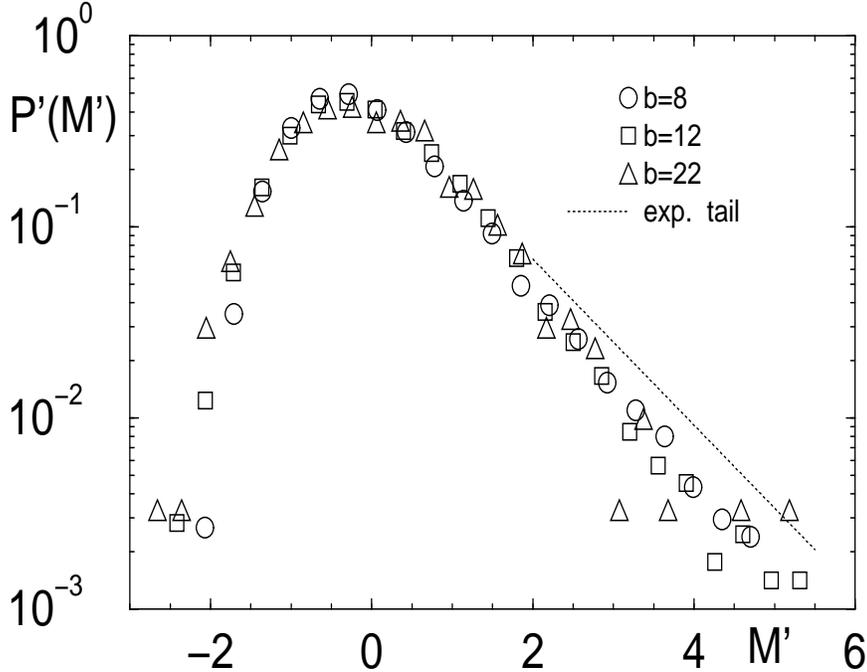,height=3.5in,width=4.5in}
\caption{Scaled probability distribution of $M'=(M-M_{av})/\sigma$. The
dotted line has been drawn to guide the eye to the exponential
(non-Gaussian) behavior.}
\label{fig5}
\end{center}
\end{figure}

In order to find further support for this anomalous behaviour, we have 
investigated the probability distribution $P(M)$ for the second 
moment at the time $\tau = 30$ (the longest time we
reached for the larger $b$-values), i.e. when the wave-function has 
almost entered the steady-state regime. The construction of reliable 
histograms has forced us to consider smaller values of $b$. In fact, 
we have studied the cases
$b=8$, $b=12$ and $b=22$, using $10^4$ realizations of the disorder 
in the first two cases and $10^3$ in the last one. The results are 
reported in Fig.~\ref{fig5}, where, following a method suggested
in Ref.~\cite{Hol98}, we have conveniently rescaled the probability 
distribution $P(M)$. In particular, defining by $M_{av}$ the average
value of $M$ at $\tau=30$ and $\sigma(M)$ the standard deviation over
the ensemble of disorder realizations, we have plotted 
$P'(M')= \sigma(M) P(M)$ vs. $M'=(M-M_{av})/\sigma$. 
After this rescaling, the distributions $P'(M')$, corresponding to the 
three $b$ values, have zero average and unit standard deviation. It 
is remarkable to notice that all curves nicely overlap indicating a 
striking scaling behaviour. A further important feature is the deviation 
from a Gaussian behaviour, especially for large values of $M'$, where a clear 
exponential tail is visible. The dotted line just above the three 
curves (corresponding to the pure exponential $\exp (-M')$) has been added
to give an idea of the decay rate which is slightly larger than 1. 
The results of this analysis are important in two respects:
i) the exponential tail `explains' the difficulties encountered in getting 
rid of statistical fluctuations in the estimate of $\Delta_M$; ii)
the deviations from a Gaussian behaviour provide an independent evidence
of the anomalous scaling behaviour of the fluctuations.
It is interesting to remark that a preliminary quantitative comparison 
has revealed a striking identity of the probability $P'(M')$ with the
distributions found in Ref.~\cite{Hol98} for
such diverse quantities as the magnetization in the 2D XY model and
the power consumption in a confined turbulent flow. This close
correspondence deserves further investigations.

\begin{figure}[hbt]
\begin{center}
\psfig{figure=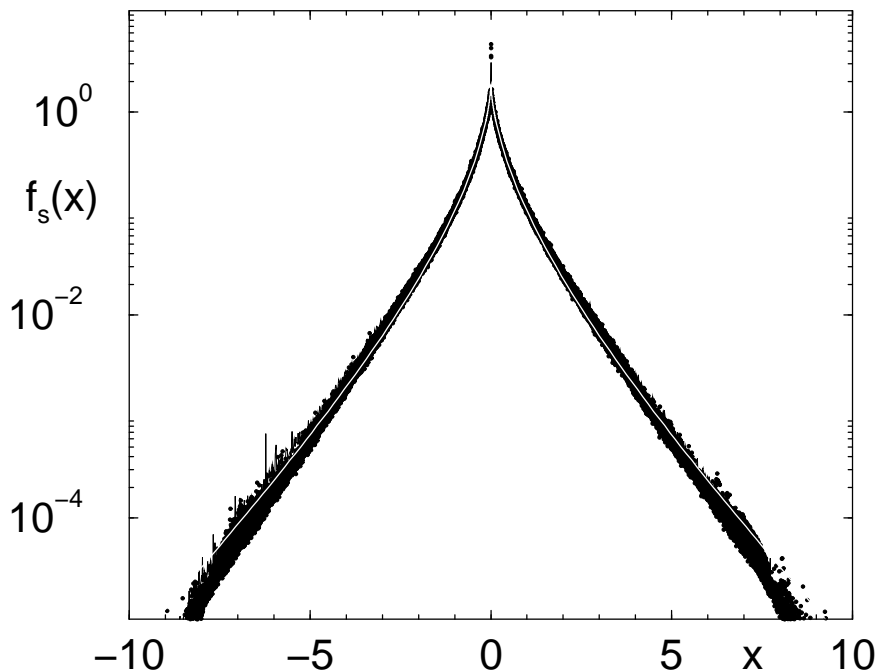,height=3.5in,width=4.5in}
\caption{Probability profile rescaled using formula (\ref{prof})
for several $b$ values vs. $x=j/b^2$, compared with Zhirov's 
theoretical prediction (white line).}
\label{fig6}
\end{center}
\end{figure}

Finally, we want to compare our results with the theoretical predictions
for the asymptotic shape of the wave-packet. In~\cite{Izr97} a reasonable
agreement was found between the numerical data and the formula obtained
by Gogolin for strictly one-dimensional systems~\cite{Gogo76}. Since a
theoretical expression has been derived in the meantime for quasi-1D systems
\cite{Zhi97}, it is desirable to compare our data also with this expression. 
In Fig.~\ref{fig6} we present the disorder-averaged probability profiles
$ \langle |\psi_j(t)|^2 \rangle_{d} = \tilde{f}(j,t)$ for large times,
rescaled under the assumption~\cite{Izr96}
\begin{equation}
f_s(x) = b^2 \tilde{f}(j,\infty) ; \qquad x=j/b^2,
\label{prof}
\end{equation}
and compare them with Zhirov's theoretical formula, which is denoted 
by the white line.  No appreciable deviation is
noticeable except for the extreme part of the tails, where it is reasonable 
to expect numerical errors due to boundary effects. 

\begin{figure}[hbt]
\begin{center}
\psfig{figure=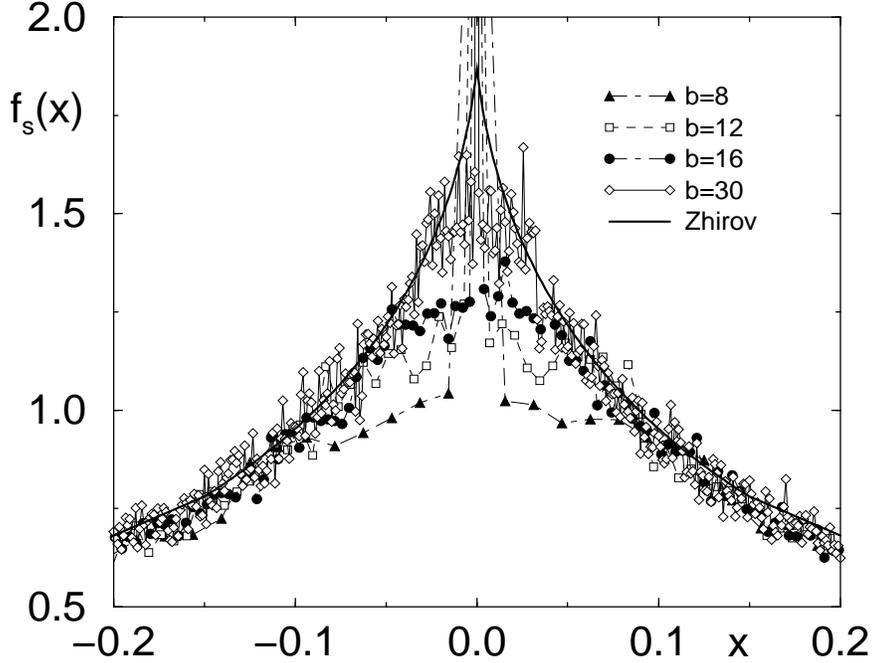,height=3.5in,width=4.5in}
\caption{Zoom of the central region of Fig.~\ref{fig6}.
Zhirov's formula is now the black line.}
\label{fig7}
\end{center}
\end{figure}

The good overlap is partly due to the (unavoidable) choice of logarithmic 
scales in Fig.~\ref{fig6}. However, if we zoom the region around the maximum
(with the exception of the zero channel), one can see
in Fig.~\ref{fig7} a slow tendency of the various curves to grow towards the 
theoretical expectation. This is consistent with the behaviour of $M$
reported in Fig.~\ref{fig1}, which reveals a convergence from above for the 
mean squared displacement.
It is interesting to notice that all such deviations are mainly due to
the finiteness of $b$ while the lack of asymptoticity in $t$ appears to
be much less relevant.

\begin{figure}[hbt]
\begin{center}
\psfig{figure=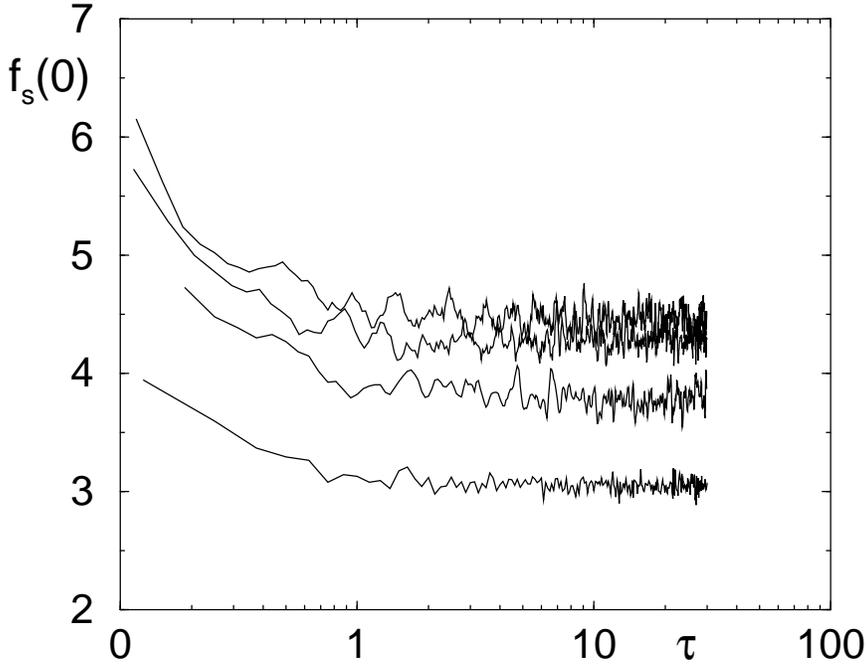,height=3.5in,width=4.5in}
\caption{Return probability to the origin $f_s(0)$ vs. $\tau$
for several $b$ values: from bottom to top $b=8,16,22,30$.}
\label{fig8}
\end{center}
\end{figure}

Finally, we consider separately the zero channel, i.e.  the return 
probability to the origin $f_s(0)$. In Fig.~\ref{fig8} we plot 
this quantity versus $\tau$ for different values of
$b$. In all cases, a quite fast convergence, as compared with 
the behaviour of the packet-width, to the asymptotic value is clearly
seen. In practice, as soon as $\tau$ is about 1, the average value of 
$f_s(0)$ reaches the asymptotic value. It is instructive to compare our 
numerical findings with the asymptotic (in time and $b$) analytic 
expression $f_s(0) =6$. By fitting the dependence of the time average of 
$f_s$ (in the interval $1<\tau<30$) on $b$ as in Eq.~\ref{fit1}), we
find that the asymptotic value is about 5.7 and that the convergence
rate is $1/\sqrt{b}$. The numerical value is in a reasonable agreement
with the theoretical one, considering that it is the result of an 
extrapolation of data already affected by errors of the order of a 
few percent. The non trivial part of the result is the rate of convergence 
of this probability, which is definitely slower than the $1/b$ behaviour 
displayed by the second (and other low order) moments.
The behaviour of the return probability, however, is in agreement with the
theoretical predictions made in~\cite{Zhi97},
where the finite $b$ deviations from the asymptotic steady-state
probability distribution were estimated to be of order
$O(1/\sqrt{b})$ in the $|x| \le 1/b$ neighbourhood of the origin. 

\begin{figure}[hbt]
\begin{center}
\psfig{figure=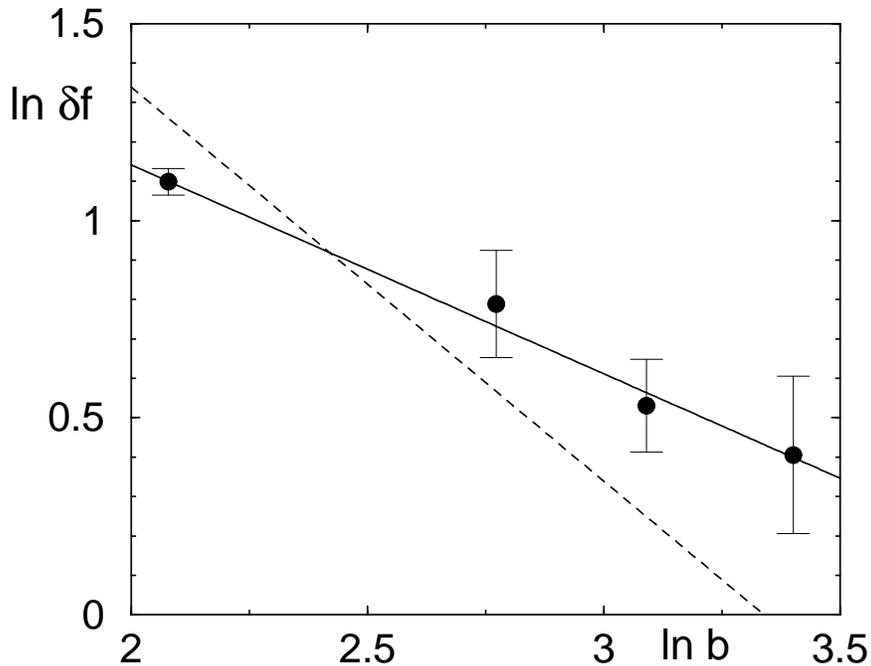,height=3.5in,width=4.5in}
\caption{Deviations of the return probability from the asymtpotic value
vs. $b$ (log-log scale). The circles represent numerical data, the dashed
line is a fit with a power law $1/b$, the continuous line is a fit with a
power law $1/b^{\alpha}$, with $\alpha = 0.53$.}
\label{fig9}
\end{center}
\end{figure}

The $b$-dependence of the return probability is further illustrated in
Fig.~\ref{fig9}, where we plot the deviation $\delta f = 6 - f_{s}(0)$
from the asymptotic value $\lim_{b \rightarrow \infty} f_{s}(0) = 6$
as a function of $b$ in bilogarithmic scale.
The displayed numerical values were obtained by averaging the
return probability both over disorder realizations and the time interval
$20 < \tau < 30$; the data were then fitted with two expressions,
exhibiting deviations from the asymptotic value $f_s(0)=6$ of order
$O(1/b)$ and $O(1/b^{\alpha})$ respectively. In the second case, the
exponent $\alpha$ was used as a fitting parameter and the best fit value
was $\alpha = 0.53$. 
As can be seen from Fig.~\ref{fig9}, the power law with $O(1/\sqrt{b})$
corrections fits the data much better than the one with deviations of
order $O(1/b)$.

\section{Conclusions and perspectives}

We have studied the time evolution of an initial $\delta$-like wave-packet 
in a 1D disordered lattice with long-range hopping.
The main results of this paper are the following.
We have confirmed with clean numerics the scaling law~(\ref{scal})
for the mean square displacement $\tilde M$, first proposed and studied 
in Ref.~\cite{Izr97}. This scaling law is valid in the large $b$ limit;
here we have found that finite $b$ corrections are of the order $1/b$.
We have proposed formula (\ref{fit2}) for fitting the time evolution of 
$\tilde M$ towards its steady state value; this formula 
contains the logarithmic corrections suggested by the existence of Mott states.
We confirm the presence of an anomaly in the scaling law of the 
relative fluctuations $\Delta_{\tilde{M}}/\tilde M$ of the mean 
square displacement, finding that they vanish for large $b$ as
$b^{-0.75}$.
We have linked this anomaly to the presence of non-Gaussian fluctuations
of the mean square displacement. In fact, the probability distribution
of $\tilde M$ displays an exponential tail for large values of
$\tilde M$. The conveniently rescaled probability strikingly coincides
with the distributions obtained in Ref.~\cite{Hol98} for
such diverse quantities as the magnetization in the 2D XY model and
the power consumption in a confined turbulent flow.
The degree of universality of such distribution deserves further
investigations.
Finally, we have compared the numerical results on the steady state
probability profile with the theoretical formula proposed by Zhirov for
large $b$, finding a good agreement. We have computed for the first time
finite $b$ corrections, obtaining $O(1/b)$ deviations for the moments of
the probability profile and $O(1/\sqrt{b})$ corrections for the return
probability to the origin.

\section{Acknowledgements}
We thank B. Chirikov for having suggested the numerical verification
of Zhirov's formula. We thank O. Zhirov for having exchanged with us
his ideas and for his comments on our numerical results.
We finally thank F.M. Izrailev, H. Kantz and B. Mehlig for 
useful discussions.


\begin{thebibliography}{10000}
\bibitem{xxx} We are indebted to an anonymous referee for this observation.
\end{thebibliography}
\end{document}